\documentclass{article}

\usepackage{microtype}
\usepackage{graphicx}
\usepackage{subfigure}
\usepackage{booktabs} 
\usepackage{url}
\usepackage{amsmath}
\usepackage{amssymb}
\usepackage{amsfonts}

\usepackage{tikz}
\usetikzlibrary{arrows,positioning,fit} 
\tikzset{
    >=stealth',
    punkt/.style={
           rectangle,
           rounded corners,
           draw=black, very thick,
           text width=6.5em,
           minimum height=2em,
           text centered},
    pil/.style={
           ->,
           thick,
           shorten <=2pt,
           shorten >=2pt,},
    lip/.style={
           <-,
           thick,
           shorten <=2pt,
           shorten >=2pt,}
}

\usepackage{hyperref}

\usepackage[accepted]{icml2021}

\icmltitlerunning{Distinctions between the CATE and ITE under Ignorability Assumptions}

\newcommand\ci{\perp\!\!\!\perp}

\begin{document}

\twocolumn[
\icmltitle{On the Distinction Between ``Conditional Average Treatment Effects'' (CATE) and ``Individual Treatment Effects'' (ITE) Under Ignorability Assumptions}



\icmlsetsymbol{equal}{*}

\begin{icmlauthorlist}
\icmlauthor{Brian G. Vegetabile}{rand}
\end{icmlauthorlist}

\icmlaffiliation{rand}{RAND Corporation, Santa Monica, CA}

\icmlcorrespondingauthor{Brian G. Vegetabile}{bvegetab@rand.org}
\icmlkeywords{Machine Learning, ICML}

\vskip 0.3in
]



\printAffiliationsAndNotice{}  

\begin{abstract}
Recent years have seen a swell in methods that focus on estimating ``individual treatment effects''. These methods are often focused on the estimation of heterogeneous treatment effects under ignorability assumptions.  This paper hopes to draw attention to the fact that there is nothing necessarily ``individual'' about such effects  under ignorability assumptions and isolating individual effects may require additional assumptions.  Such individual effects, more often than not, are more precisely described as ``conditional average treatment effects'' and confusion between the two has the potential to hinder advances in personalized and individualized effect estimation.  
\end{abstract}

\section{Introduction} \label{sec:intro}

Achieving personalized, or individualized, effect estimation is an ambitious goal throughout science.  Further, the capability to estimate individual effects within populations that generalize to new settings would dramatically alter how we approach medicine (and potentially many other disciplines).  To that end, there has been a growing body of literature within the field of causal inference on the estimation of heterogeneous treatment effects \cite{bart2011,kennedy2020optimal} and on optimal learning of individualized treatment policies \cite{murphy2003optimal,kallus2021minimax}.  Similarly, experimental designs such as case-crossover designs \cite{maclure1991case,marshall1993analysis} have long been used in an attempt to estimate ``individual'' effects and limit the effects of within-individual variability through experimental design.  

Recently though, there has been growing usage of the term ``individual treatment effects'' (ITE) to describe methods that focus on exploiting heterogeneity among effects within a population \cite{pmlr-v70-shalit17a,lu2018estimating} and to nonparametrically estimate effects for individuals conditioned on observed covariates.  While these methods all estimate effects conditioned on covariates and thus can be useful for personalizing medicine, the strong assumptions that they employ do not necessarily imply that these are the effects for the \textit{individual}. In fact, the true effect for an individual may actually differ in both magnitude and direction from those estimated using the approaches described.  The goal of this paper is to make clear the distinctions between individual treatment effects (ITE) and conditional average treatment effects (CATE) when strong ignorability assumptions are made.  

The arguments made here all relate to the fact that the strong ignorability assumptions \cite{imbens_rubin_2015} employed only guarantee that, given certain covariates, it is possible to \textit{ignore} other covariates as if they were randomized.  In Section \ref{sec:notation}, notation and general assumptions are described, including a distinction on the differences between ``individual effect'' estimands and ``conditional effect'' estimands.  Section \ref{sec:bias} provides heuristic arguments on the importance of the ignorability assumptions and demonstrates how unobserved covariates (but potentially important variables) can be ignored in an analysis provided the assumptions are met. Section \ref{sec:linearmodelsdemo} provides a simple example, where in the presence of interactions among an exposure and an unobserved covariate, individual effects are not guaranteed to have even the same sign as conditional average effects. Finally,  Section \ref{sec:rct} discusses the strong ignorability assumption in context of randomized controlled trials and implications for CATE estimation in this setting. 

The hope is that this paper illustrates some of the difficulties in estimating individual effects and how ignorability assumptions and the estimation of conditional effects are not sufficient for individual inference.  

\section{Notation \& Assumptions: Causal Inference Framework} \label{sec:notation}

 Let $A$ be a binary random variable representing treatment assignment and let $Y$ be a random variable that represents the observed outcome. Let $L=(X,Z)$ be a vector of pre-exposure covariates\footnote{Note: $L=(X,Z)$ could also be considered $L=(X^{obs},X^{mis})$, but the notation above is meant to align with the literature and convey the distinction that $X$ are the set that provide ignorability and $Z$ are hidden/unobserved variables} with length $p=d+q$ where $X$ is a vector of length $d$ of observed covariates (i.e.,  collected and available in a data set) and $Z$ is a vector of of length $q$ of unobserved covariates (i.e., $Z$ is not measured and $q$ may be large). 
 
\begin{figure}[H]
        \centering
        \begin{tikzpicture}[node distance=0.25cm, auto,]
            \node[] (Y) {$Y$};
            \node[left= of Y] (d1) {};
            \node[left= of d1] (A) {$A$}
                edge[pil] (Y.west);
            \node[left= of A] (d2) {};
            \node[draw=black,left= of d2] (X) {$X$}
                edge[pil, bend left=45] (Y.north west)
                edge[pil] (A.west);
            \node[draw=black,below= of X] (Z) {$Z$}
                edge[thick] (X.south)
                edge[pil, bend right=20] (Y.south);    
            
            \node[dotted, draw=black,fit=(X) (Z), inner sep=0.2cm] (boxes) {};
        \end{tikzpicture}
        \caption{Notional Graph Describing the Relationships Among the Variables.  Only $X$ is used to assign exposure levels and $Z$ is unobserved but related to the outcome variable $Y$.}
        \label{fig:dag}
    \end{figure}
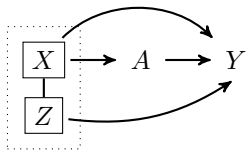

 The variables will be assumed to loosely follow the graph in Figure \ref{fig:dag} where the set $L$ temporally occurs before $A$ (and $X$ and $Z$ can be correlated, or exhibit some other stronger causal structure) and within $L$ only $X$ is used to assign treatment levels. We note that while $Z$ is unobserved, based on the representation in this graph, this set of covariates does not necessarily represent \textit{hidden confounding}, in that, given $X$, the distribution of $Z|A,X$ is equivalent to $Z|X$.    

 Throughout we will focus on defining effects in the Neyman/Rubin Potential Outcome Framework (see Imbens \& Rubin, 2015, for an introduction) where the potential outcome for each unit under each treatment $A=a$ is defined as $Y(a) \ \forall \ a\in\{0,1\}$.  Under this framework it is common to define an \textit{individual's treatment effect} as a contrast between potential outcomes, e.g., 
 \begin{align}
     \tau_i = Y_i(1) - Y_i(0)
 \end{align}
 Due to the \textit{Fundamental Problem of Causal Inference} \cite{holland1986} we can only observe one potential outcome for each individual and thus the framework typically considers how \textit{average} effects can be identified such as the following estimands,
 \begin{align*}
     \tau(x) &= E[Y(1) - Y(0) \ | \ X = x] && \mbox{(CATE)} \\
     E_X[\tau(x)] &= E_X[E[Y(1) - Y(0) \ | \ X = x]] && \mbox{(ATE)}
 \end{align*}
 where $\tau(x)$ is the \textit{conditional average treatment effect} among individuals with the same vector of covariates $X$ and $E_X[\tau(x)]$ is the \textit{average treatment effect} over a population represented by the distribution of $X$ \cite{li2018balancing}. The above should make the clear distinction that $\tau_i$ does not necessarily equal $\tau(x)$, where the first is an individual's effect and the second is an average among the population.   

Identification of the CATE and ATE within the Neyman/Rubin Potential Outcome Framework (in its simplest form) generally require the following sets of assumptions: 1) ``Strong Ignorability'' given a set of covariates (in this case $X$), and 2) the Stable Unit Treatment Value Assumption (SUTVA). These are listed below:

\paragraph{Assumptions 1a \& 1b: Strong Ignorability Given a Set of Covariates \cite{rosenbaum_rubin_1983} -} We say that a treatment assignment mechanism is strongly ignorable given a set of covariates $X$ if $Y(1),Y(0) \ci A \ | \ X$ and $0 < \mbox{Pr}(A = 1 | X = x)< 1$ for any $x$ of interest.  

The first part requires conditional independence between the treatment and the potential outcomes given the set of covariates $X$ and the second part of the assumption is often referred to the \textit{positivity} assumption.  The positivity assumption typically is evaluated using the function $e(x) = \mbox{Pr}(A = 1 | X = x)$ referred to as the propensity score \cite{rosenbaum_rubin_1983}.

\paragraph{Assumption 2a \& 2b: SUTVA  - \cite{rubin1980, imbens_rubin_2015}} The treatment assignment of one unit does not affect the potential outcomes of another unit (i.e., \textit{no interference} among units) and no hidden variability in treatment levels (e.g., each 100mg aspirin tablet is equivalent). 

Assumptions 2a and 2b generally are generally used to imply consistency between the potential outcomes and the observed outcomes such that we can assert that $Y_i(a) = \sum_{a} I(A_i = a) Y_i$ enabling estimation of effects from the outcomes.  

There are many results that demonstrate that if SUTVA and strong ignorability given $X$ holds, i.e.,  $Y(1),Y(0) \ci A \ | \ X$, then both the CATE and ATE are identified and estimable \cite{imbens_rubin_2015,pearl2016causal}.

\section{On the Ignorability Assumption and Bias} \label{sec:bias}

To understand the distinction between an ITE and a CATE, it is important to first understand the role of the ignorability assumption in reducing bias in studies. In this section we provide a heuristic discussion on the ignorability assumption and how it enables estimation, but does not provide a guarantee that estimation is individualistic. What follows is largely similar to the arguments of Cochran \& Rubin \cite{cochranrubin1973}. 

Consider the following simple relationship,
\begin{align*}
    Y = \beta_0 + \beta_1 A + \beta_2 X + \beta_3 Z
\end{align*}
The true effect in the population (i.e., the ATE) is $\beta_1$, but naive estimation of $\beta_1$ from a model of $E[Y|A]$ may be biased based on the structure present in Figure \ref{fig:dag}.  

Specifically, bias could enter estimation because of potential differences in the covariate distributions between the different treated subpopulations, e.g., differences between $E[X|A=1]$ and $E[X|A=0]$. That is, we have that
\begin{align*}
    & E[Y|A=1] - E[Y|A=0]  \\
    &= \beta_1 + \beta_2 (E[X|A=1] - E[X|A=0]) \\
    &\hspace{1em}+ \beta_3 (E[Z|A=1] - E[Z|A=0]).
\end{align*}

Bias could be removed from estimates above in two ways: 1) conditioning on \textit{all} variables important to the outcome and estimating the response function directly and using this to estimate the difference, or 2) finding a set of variables $X$ such that once conditioned upon, it follows that $E[Z|A=a, X=x] = E[Z|X=x]$ and we can \textit{ignore} the previous differences between the groups within this variable to integrate over it (and this example the remaining differences, i.e., $E[Z|A=1, X=x] - E[Z|A=0, Xx=x]$, will all become zero).  It does not mean they are not important in the outcome model regression, just that in the contrast we are estimating they can be ultimately ignored.

For example, if we assume ignorability and condition on $X$ and $A$, we have 
\begin{align*}
    & E[Y|A=1,X=x] - E[Y|A=0,X=x]  \\
    &= \beta_1 + \beta_2 (E[X|A=1,X=x] - E[X|A=0,X=x]) \\
    &\hspace{1em}+ \beta_3 (E[Z|A=1,X=x] - E[Z|A=0,X=x]) \\
    &= \beta_1 + \beta_3 (E[Z|X=x] - E[Z|X=x]) = \beta_1
\end{align*}
and here bias is no longer a function of the unobserved covariates. Additionally, we can now integrate this over the distribution of $X$, i.e., $\tau = E_X\big[E[Y|A=1,X=x] - E[Y|A=0,X=x]\big]$, to get the population treatment effect.  

Thus, one of the largest benefits of the ignorability assumption is that it would allow researchers to potentially remove bias in estimated treatment effects that could arise from variables that are important for predicting the outcome, but are not possible to control for.

\section{A Demonstration With Linear Models} \label{sec:linearmodelsdemo}

One clear example where a CATE will diverge from an ITE is in a linear model when there is an interaction between the exposure $A$ and the unobserved variable $Z$. To demonstrate the implications of the ignorability assumption and our ability to estimate the CATE and ITE, a simple simulation based on a model with such interactions is used to further illustrate the distinction between the two estimands.  

The focus again is illustrating how ignorability provides consistent estimation for the CATE while providing an ability to ignore important variables in the outcome relationship, but there may remain important differences between the estimated CATE and the true ITE.  

Let, 
\begin{align*}
    \left(\begin{array}{c} X_i \\ Z_i  \end{array}\right) \sim \mathcal{N}\bigg(\left(\begin{array}{c} 0 \\ 0  \end{array}\right), \left(\begin{array}{cc} 1 & 3\rho \\ 3\rho & 3^2  \end{array}\right)\bigg)
\end{align*}
and
$A_i\sim Bernoulli(p_A)$ with $p_A = 1 / (1  + \exp(-X_i))$, and independent error terms $\epsilon_i \sim \mathcal{N}(0,1)$.  Let the observed outcome variable be defined as 
\begin{align*}
    Y_i = 3 + A_i + A_i(X_i + Z_i) + \epsilon_i .
\end{align*}
In this process, it is assumed that $Z$ is unobserved and unavailable in the analysis.  But, if $X$ is observed and conditioned on, then it is possible to satisfy the strong ignorability assumptions required to estimate the CATE, $\tau(x)$.

Under consistency assumptions, where $Y_i(a) = \sum_a I(A_i=a) Y_i$, we can analytically derive both the ITE and the CATE, where an individual specific effect would be,
\begin{align*}
    \tau_i = 1 + x_i + z_i 
\end{align*}
and the conditional average effect in the population would be 
\begin{align*}
    \tau(x) &= 1 + x_i + E[Z_i | X_i = x_i]
    \\&= 1 + x_i (1 + 3\rho). 
\end{align*}
due to the fact that $E[Z|X=x] = \mu_z + \rho \frac{\sigma_z}{\sigma_x}(x - \mu_x)$ under properties of Gaussian distributions. 

Note that in this example we have satisfied strong ignorability, but the conditional average effect is not equal to the individual effect, i.e., $\tau(x) \ne \tau_i$.  Further, when $\rho = -1/3$, it is possible to have a situation where there is a constant CATE (i.e., $\tau(x) = 1 \  \forall \ x$), but the individual effect will still vary and could be negative for many individuals due to the large variability in $Z$. In the appendix, it is shown that for a general form of this model and data-generating process, that while it is possible that there is no correlation between the CATE and ITE, the ITE and CATE should generally have positive correlation.  

We can consistently estimate the CATE by specifying a regression model of the form
\begin{align*}
    E[Y|A,X] = \beta_0 + \beta_1 A + \beta_2 X + \beta_3 AX,
\end{align*}
where $Z$ is not included because it is not observed. Under this model, it would follow that an estimate for the CATE would be $\hat\tau(x) = \hat\beta_1 + \hat\beta_3 x$.  Strong ignorabliity implies that we will have consistent estimation for each $\hat \beta_j$. 

\subsection{Simulation Results}

To illustrate the point that consistent estimation is possible even in the presence of $Z$ in the causal structure, the data generating process of the previous section was conducted across $1000$ replications and each replication contained $N=2500$ observations. We present results for two settings of $\rho \in \{-0.3, 0.7\}$. 

\begin{figure}[H]
    \centering
    \includegraphics[width=0.8\linewidth]{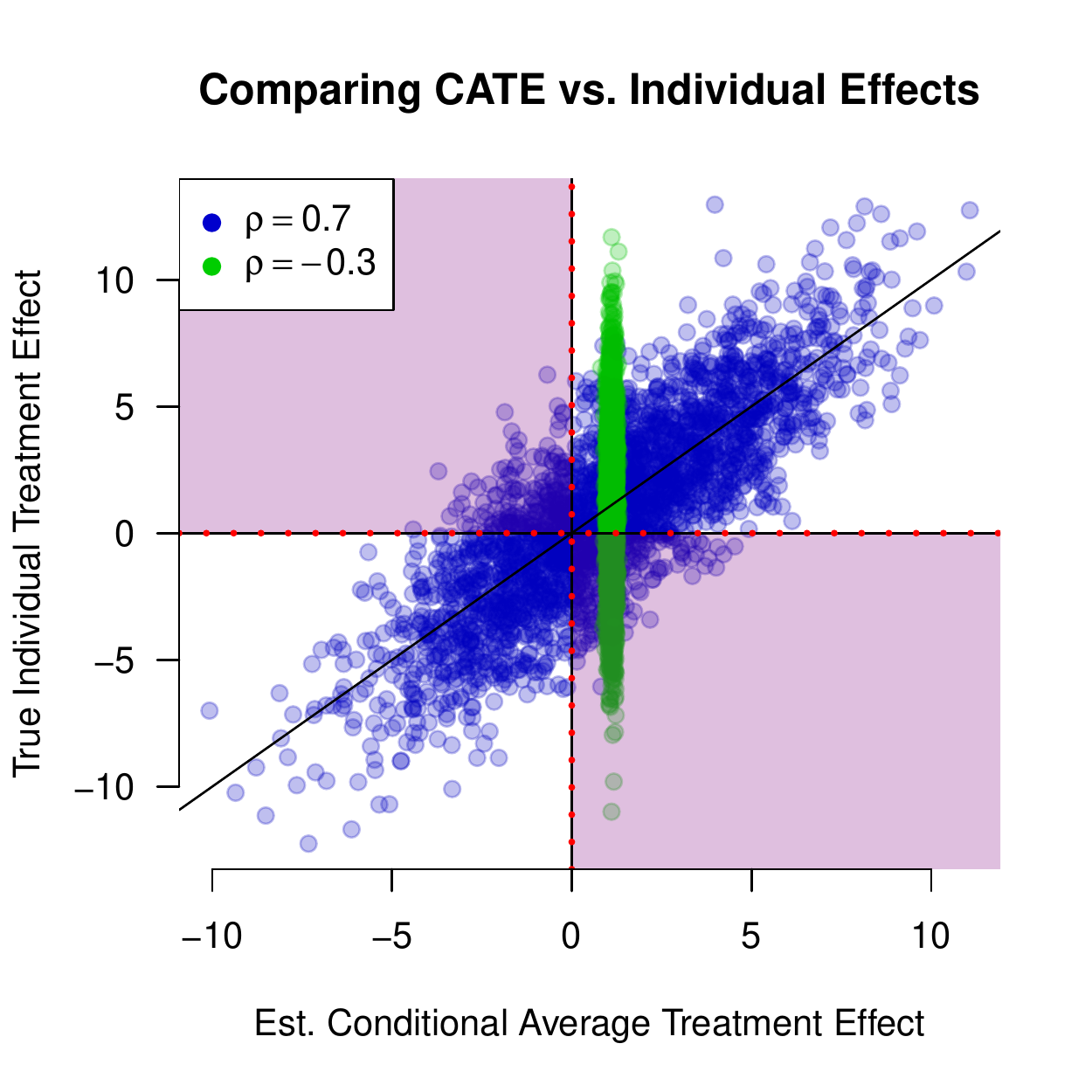}
    \caption{Comparing the estimated CATE to the true ITE for two different data generating processes described in Section \ref{sec:linearmodelsdemo}.  The diagonal  represents the 45$^o$ line and estimates along this line would occur when the CATE is equal to the ITE. The blue dots represent when the correlation between $X$ and $Z$ is high, i.e., $\rho = 0.7$.  The green dots represents a case where the correlation between $X$ and $Z$ completely removes the heterogeneity in the estimated CATE, while large variability remains in the ITE.  The red sections represent areas where there is disagreement between the CATE and ITE.}
    \label{fig:correlation}
\end{figure}

The average MSE of the CATE was 0.013 and 0.020 respectively, indicating good performance in estimating the CATE under ignorability assumptions. Figure \ref{fig:correlation} illustrates the point that an individual's true effect may differ in both magnitude and direction from an estimated CATE. Though in the case when $\rho=0.7$, the high correlation between $X$ and $Z$ can make estimates of the CATE closer to the ITE.  We see though when $\rho = -0.3$ the estimated CATE is almost a constant while there is a large variability in the true ITE.  

The case when $\rho=-0.3$ is compelling in that there can be no heterogeneity in the CATE, but large variability in individual effects.

\subsection{What if the Unobserved Variable is Not Ignorable?}

To demonstrate the distinction of the effect of $Z$ between cases when it is ignorable and when it confounds relationships, this section presents simulation results when $Z$ is directly related to $A$ in the assignment mechanism, i.e., the propensity score is a function of $z$, i.e., $e(x,z)=\mbox{Pr}(A=1|X=x,Z=z)$. To induce unobserved confounding, we change the assignment mechanism to be $p_A = 1 / (1  + \exp(Z_i - X_i))$ and repeat the simulation with $\rho = 0.7$.  

\begin{figure}
    \centering
    \includegraphics[width=0.8\linewidth]{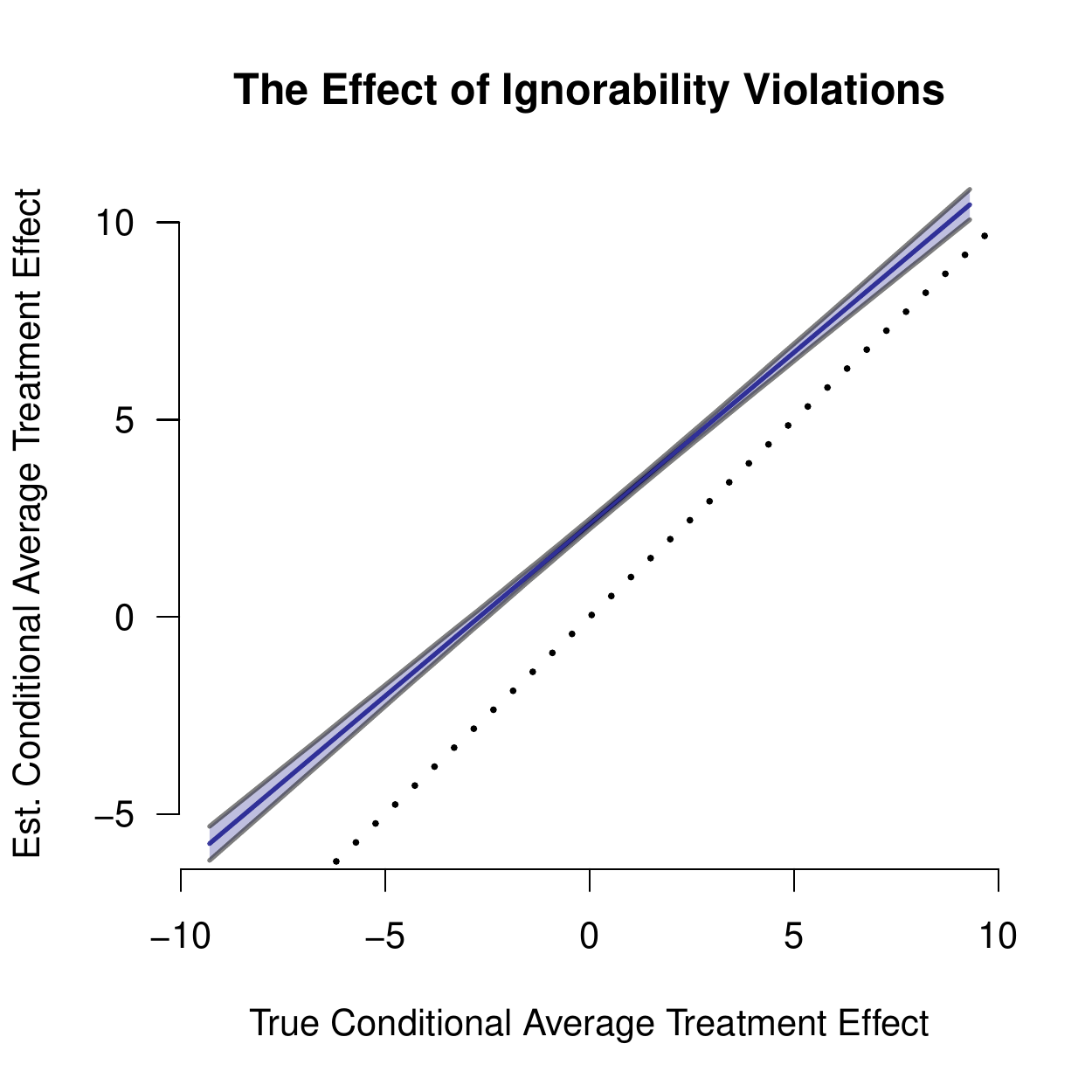}
    \caption{Comparing estimates of the CATE across simulations with the true conditional average effects when the ignorability assumption is violated. The blue region and black lines represents the mean estimates and 95\% error band on a grid of points. The dotted line represents the 45$^o$ line and if the error band contained the dotted line would represent agreement between estimates and truth. }
    \label{fig:catevtrueagain}
\end{figure}

The results indicate that what should be expected that results are now biased.  The average MSE of the CATE across replications rises to 5.699 (as compared with 0.013 earlier) and Figure \ref{fig:catevtrueagain} demonstrates the extent of the bias. This demonstrates that even knowing the correct form for the CATE, it is possible to have biased estimation if the ignorability assumption is violated (i.e., there is unobserved confounding). 

\section{RCTs, Ignorability, and CATE estimation} \label{sec:rct}

The previous sections clearly demonstrate the distinctions between the CATE and ITE in observational studies, but a more nuanced distinction about the CATE is that under a completely randomized experiment there are many different estimable CATEs (with potentially many contradictory trends).  

For example, consider a completely randomized experiment where the probability of assignment to either treatment is a constant.  Further consider a partition of the observed covariates $X = (X_{S}, X_{\neg S})$  Under this design, the strong ignorability assumption is satisfied given both sets $X_{S}$ and $X_{\neg S}$ and thus we are able to identify and consistently estimate both,
\begin{align*}
     \tau(\tilde{x}_{S}) &= E[Y(1) - Y(0) \ | \ X_{S} = \tilde{x}_{S}] && \mbox{and}\\
     \tau(\tilde{x}_{\neg S}) &= E[Y(1) - Y(0) \ | \ X_{\neg S} = \tilde{x}_{\neg S}].
\end{align*}
More specifically, for any $X^* \subseteq X$ we would also have identification for 
\begin{align*}
     \tau(x^*) &= E[Y(1) - Y(0) \ | \ X^* = x^*]
\end{align*}
Illustrating that there are a multitude of potential CATEs that might be of interest to study. 

To further this point, consider the following outcome function,
\begin{align*}
    Y = A(X_1^2 - X_2^2) + A Z^2 + A
\end{align*}
where $A$ is binary and assigned completely at random and where $X_j\sim\mathcal{N}(0,1)$, $Z\sim\mathcal{N}(0,1)$, with $X_j\perp Z$ and $X_j\perp X_{j'}$. It follows that the individual effect would be
\begin{align*}
    \tau_i = x_{i1}^2 - x_{i2}^2 + z_i^2 + 1
\end{align*}
and one version of a CATE that is identified is 
\begin{align*}
    E[Y(1) - Y(0) | X_1 = x_1] = x_1^2 +1
\end{align*}
and another is
\begin{align*}
    E[Y(1) - Y(0) | X_2 = x_2] = 3 - x_2^2 
\end{align*}
demonstrating that it is possible to have two distinct CATES, both with complex contradictory relationships (one a positive quadratic and the other a negative quadratic).  Finally, if $Z$ were not observed, even in a randomized controlled trial using only an ignorability assumption, again it would not be possible to estimate the ITE.  

Estimation of an ITE is complicated by both a need to get the functional form correct, but also a need to collect the right set of variables in the data.  

\section{Implications} \label{sec:imps}

The discussion here does not diminish the importance of individual treatment effect estimation, it just implies that ignorability alone is not sufficient for the estimation of individual effects; further the point should be made that the CATE is not an ITE, though both may be correlated. Both are important estimands and in many cases the scientific question of interest may actually call for precise knowledge of the CATE. Further work should investigate the ability to either analytically bound an estimate of the ITE from the CATE, or work on the development of sensitivity analyses that may be helpful to understand the range of reasonable values for the ITE given an estimated CATE.

Fully establishing identification for the estimation of individual effects would often require additional assumptions (potentially very strong assumptions) such as those in case-crossover designs or experiments that contain within-individual repeated observations \cite{murphy2003optimal}.  In many cases, individual estimation requires a willingness to make these strong assumptions about the suitability for observations to serve as counterfactuals.

It is important to note that while the best way to satisfy ignorability is through design, simple designs alone are not enough to imply that we have the power to precisely estimate individual effects. Finally, when ignorability is asserted in observational studies it should be done so judiciously and with caveats that accurately capture where inference could be applicable. 

\section*{Acknowledgements}

The author would like to thank the many researchers that helped to think critically about these ideas as well as the workshop reviewers for their feedback on this draft. In particular, Matthew Cefalu, Daniel McCaffrey, Beth Ann Griffin and Daniel Gillen for their advice and support in preparing this work.

Research reported in this manuscript was supported by NIDA of the National Institutes of Health under award number R01DA045049.  The content is solely the responsibility of the authors and does not necessarily represent the official views of the National Institutes of Health.

\bibliography{cate_v_ite_paper}

\begin{thebibliography}{15}
\providecommand{\natexlab}[1]{#1}
\providecommand{\url}[1]{\texttt{#1}}
\expandafter\ifx\csname urlstyle\endcsname\relax
  \providecommand{\doi}[1]{doi: #1}\else
  \providecommand{\doi}{doi: \begingroup \urlstyle{rm}\Url}\fi

\bibitem[Cochran \& Rubin(1973)Cochran and Rubin]{cochranrubin1973}
Cochran, W.~G. and Rubin, D.~B.
\newblock Controlling bias in observational studies: A review.
\newblock \emph{Sankhya: The Indian Journal of Statistics, Series A
  (1961-2002)}, 35\penalty0 (4):\penalty0 417--446, 1973.
\newblock ISSN 0581572X.
\newblock URL \url{http://www.jstor.org/stable/25049893}.

\bibitem[Hill(2011)]{bart2011}
Hill, J.~L.
\newblock Bayesian nonparametric modeling for causal inference.
\newblock \emph{Journal of Computational and Graphical Statistics}, 20\penalty0
  (1):\penalty0 217--240, 2011.
\newblock \doi{10.1198/jcgs.2010.08162}.
\newblock URL \url{http://dx.doi.org/10.1198/jcgs.2010.08162}.

\bibitem[Holland(1986)]{holland1986}
Holland, P.~W.
\newblock Statistics and causal inference.
\newblock \emph{Journal of the American Statistical Association}, 81\penalty0
  (396):\penalty0 945--960, 1986.
\newblock ISSN 01621459.
\newblock URL \url{http://www.jstor.org/stable/2289064}.

\bibitem[Imbens \& Rubin(2015)Imbens and Rubin]{imbens_rubin_2015}
Imbens, G.~W. and Rubin, D.~B.
\newblock \emph{Causal Inference for Statistics, Social, and Biomedical
  Sciences: An Introduction}.
\newblock Cambridge University Press, 2015.
\newblock \doi{10.1017/CBO9781139025751}.

\bibitem[Kallus \& Zhou(2021)Kallus and Zhou]{kallus2021minimax}
Kallus, N. and Zhou, A.
\newblock Minimax-optimal policy learning under unobserved confounding.
\newblock \emph{Management Science}, 67\penalty0 (5):\penalty0 2870--2890,
  2021.

\bibitem[Kennedy(2020)]{kennedy2020optimal}
Kennedy, E.~H.
\newblock Optimal doubly robust estimation of heterogeneous causal effects.
\newblock \emph{arXiv preprint arXiv:2004.14497}, 2020.

\bibitem[Li et~al.(2018)Li, Morgan, and Zaslavsky]{li2018balancing}
Li, F., Morgan, K.~L., and Zaslavsky, A.~M.
\newblock Balancing covariates via propensity score weighting.
\newblock \emph{Journal of the American Statistical Association}, 113\penalty0
  (521):\penalty0 390--400, 2018.

\bibitem[Lu et~al.(2018)Lu, Sadiq, Feaster, and Ishwaran]{lu2018estimating}
Lu, M., Sadiq, S., Feaster, D.~J., and Ishwaran, H.
\newblock Estimating individual treatment effect in observational data using
  random forest methods.
\newblock \emph{Journal of Computational and Graphical Statistics}, 27\penalty0
  (1):\penalty0 209--219, 2018.

\bibitem[Maclure(1991)]{maclure1991case}
Maclure, M.
\newblock The case-crossover design: a method for studying transient effects on
  the risk of acute events.
\newblock \emph{American journal of epidemiology}, 133\penalty0 (2):\penalty0
  144--153, 1991.

\bibitem[Marshall \& Jackson(1993)Marshall and Jackson]{marshall1993analysis}
Marshall, R.~J. and Jackson, R.~T.
\newblock Analysis of case-crossover designs.
\newblock \emph{Statistics in medicine}, 12\penalty0 (24):\penalty0 2333--2341,
  1993.

\bibitem[Murphy(2003)]{murphy2003optimal}
Murphy, S.~A.
\newblock Optimal dynamic treatment regimes.
\newblock \emph{Journal of the Royal Statistical Society: Series B (Statistical
  Methodology)}, 65\penalty0 (2):\penalty0 331--355, 2003.

\bibitem[Pearl et~al.(2016)Pearl, Glymour, and Jewell]{pearl2016causal}
Pearl, J., Glymour, M., and Jewell, N.~P.
\newblock \emph{Causal inference in statistics: A primer}.
\newblock John Wiley \& Sons, 2016.

\bibitem[Rosenbaum \& Rubin(1983)Rosenbaum and Rubin]{rosenbaum_rubin_1983}
Rosenbaum, P.~R. and Rubin, D.~B.
\newblock The central role of the propensity score in observational studies for
  causal effects.
\newblock \emph{Biometrika}, 70\penalty0 (1):\penalty0 41--55, 1983.
\newblock \doi{10.1093/biomet/70.1.41}.
\newblock URL \url{http://biomet.oxfordjournals.org/content/70/1/41.abstract}.

\bibitem[Rubin(1980)]{rubin1980}
Rubin, D.~B.
\newblock Comment on `randomization analysis of experimental data: The fisher
  randomization test' by d. basu.
\newblock \emph{Journal of the American Statistical Association}, 75\penalty0
  (371):\penalty0 591--593, 1980.
\newblock \doi{10.1080/01621459.1980.10477517}.
\newblock URL \url{http://dx.doi.org/10.1080/01621459.1980.10477517}.

\bibitem[Shalit et~al.(2017)Shalit, Johansson, and Sontag]{pmlr-v70-shalit17a}
Shalit, U., Johansson, F.~D., and Sontag, D.
\newblock Estimating individual treatment effect: generalization bounds and
  algorithms.
\newblock In Precup, D. and Teh, Y.~W. (eds.), \emph{Proceedings of the 34th
  International Conference on Machine Learning}, volume~70 of \emph{Proceedings
  of Machine Learning Research}, pp.\  3076--3085. PMLR, 06--11 Aug 2017.
\newblock URL \url{http://proceedings.mlr.press/v70/shalit17a.html}.

\end{thebibliography}
\bibliographystyle{icml2021}

\begin{appendix}
\section{Supplement to Section 4}

Assume the graph in Figure \ref{fig:dag} and let $X \sim \mathcal{N}(\mu_x, \sigma_x^2)$ and $Z \sim \mathcal{N}(\mu_z, \sigma_z^2)$ and assume a correlation $\rho$ between $X$ and $Z$. Further assume the following model for the data generating process for the outcome variable for each individual $i$, 
\begin{align*}
    Y_i = \beta_0 + \beta_1 A_i + \beta_2 X_i + \beta_3 Z_i + \beta_4 A_i X_i + \beta_5 A_i Z_i
\end{align*}
Under this model, the ITE for an individual with covariate values $X_i =x$ and $Z_i = z$ would be
\begin{align*}
    \tau_i = \beta_1 + \beta_4 x + \beta_5 z
\end{align*}
and the CATE would be
\begin{align*}
    \tau(x) &= \beta_1 + \beta_4 x + \beta_5 E[Z_i|X_i=x] \\ 
    &= \beta_1 + \beta_4 x + \beta_5 \left(\mu_z + \rho \frac{\sigma_z}{\sigma_x} (x - \mu_x)\right)  \\
    &= \left( \beta_1 + \beta_5 \left(\mu_z - \rho \frac{\sigma_z}{\sigma_x} \mu_x \right) \right) + \left(\beta_4 + \beta_5 \rho \frac{\sigma_z}{\sigma_x} \right) x
\end{align*}


To further investigate the relationship between the ITE and the CATE, we can investigate their covariance and correlation. In this population, it follows that the covariance between the CATE and ITE is
\begin{align*}
    & Cov(\tau(X_i) , \tau_i ) \\
    &= \beta_4 \left(\beta_4 + \beta_5 \rho \frac{\sigma_z}{\sigma_x} \right) \sigma_x^2 + \beta_5 \left(\beta_4 + \beta_5 \rho \frac{\sigma_z}{\sigma_x} \right) \rho \sigma_x \sigma_z \\ 
    &= \sigma_x^2 \left(\beta_4 + \beta_5 \rho \frac{\sigma_z}{\sigma_x} \right)^2,
\end{align*}
and that $Cov(\tau(X_i), \tau(X_i)) = \sigma_x^2 \left(\beta_4 + \beta_5 \rho \frac{\sigma_z}{\sigma_x} \right)^2$ and $Cov(\tau_i, \tau_i) = \beta_4^2 \sigma_x^2 + \beta_5^2 \sigma_z^2 + 2 \beta_4 \beta_5 \rho \sigma_x \sigma_z$. Thus,
\begin{align*}
    & Corr(\tau(X_i) , \tau_i ) \\
    &= \frac{\sigma_x^2 \left(\beta_4 + \beta_5 \rho \frac{\sigma_z}{\sigma_x} \right)^2}{\sqrt{\sigma_x^2 \left(\beta_4 + \beta_5 \rho \frac{\sigma_z}{\sigma_x} \right)^2}\sqrt{\beta_4^2 \sigma_x^2 + \beta_5^2 \sigma_z^2 + 2 \beta_4 \beta_5 \rho \sigma_x \sigma_z}} \\
    &= \frac{\left|\sigma_x \left(\beta_4 + \beta_5 \rho \frac{\sigma_z}{\sigma_x} \right)\right|}{\sqrt{\beta_4^2 \sigma_x^2 + \beta_5^2 \sigma_z^2 + 2 \beta_4 \beta_5 \rho \sigma_x \sigma_z}}.
\end{align*}
Therefore, under this outcome model and data-generating process it follows that
\begin{align*}
    Corr(\tau(X_i) , \tau_i ) \ge 0
\end{align*}
with equality when $\beta_4 + \beta_5 \rho \frac{\sigma_z}{\sigma_x}$ is equal to zero.  That is, when there is no heterogeneity with respect to $X$ in the CATE.
\end{appendix}

\end{document}